%% file: paper.tex
\documentclass[conference]{IEEEtran}
\IEEEoverridecommandlockouts

\usepackage{cite}
\usepackage{amsmath,amssymb,amsfonts}
\usepackage{algorithmic}
\usepackage{graphicx}
\usepackage{textcomp}
\usepackage{xcolor}
\usepackage[all]{nowidow}
\usepackage{url}
\usepackage{balance}
\usepackage[all]{nowidow}
\usepackage{subcaption}

\def\BibTeX{{\rm B\kern-.05em{\sc i\kern-.025em b}\kern-.08em
    T\kern-.1667em\lower.7ex\hbox{E}\kern-.125emX}}

\begin{document}

\title{Edge (of the Earth) Replication: Optimizing Content Delivery in Large LEO Satellite Communication Networks\\
    \thanks{Funded by the Deutsche Forschungsgemeinschaft (DFG, German Research Foundation) -- 415899119.}
}

\author{\IEEEauthorblockN{Tobias Pfandzelter, David Bermbach}
    \IEEEauthorblockA{\textit{Technische Universit\"at Berlin \& Einstein Center Digital Future}\\
        \textit{Mobile Cloud Computing Research Group} \\
        \{tp,db\}@mcc.tu-berlin.de}
}

\maketitle

\begin{abstract}
    Large low earth orbit (LEO) satellite networks such as SpaceX's Starlink constellation promise to deliver low-latency, high-bandwidth Internet access with global coverage.
    As an alternative to terrestrial fiber as a global Internet backbone, they could potentially serve billions of Internet-connected devices.
    Currently, operators of CDNs exploit the hierarchical topology of the Internet to place points-of-presence near users, yet this approach is no longer possible when the topology changes to a single, wide-area, converged access and backhaul network.

    In this paper, we explore the opportunities of points-of-presence for CDNs within the satellite network itself, as it could provide better access latency for users while reducing operational costs for the satellite Internet service providers.
    We propose four strategies for selecting points-of-presence in satellite constellations that we evaluate through extensive simulation.
    In one case, we find that replicating web content within satellites can reduce bandwidth usage in the constellation by 93\% over an approach without replication in the network, while storing only 0.01\% of all content in individual satellites.
\end{abstract}


\input{Sections/1_introduction}
\input{Sections/2_background}

\input{Sections/3_pop_strategies}
\input{Sections/4_simulation}
\input{Sections/5_discussion}
\input{Sections/6_related_work}
\input{Sections/7_conclusion}

\balance

\bibliographystyle{IEEEtran}
\bibliography{bibliography}

\end{document}

%% file: Sections/1_introduction.tex
\section{Introduction}
\label{sec:introduction}

Initiatives by companies such as SpaceX, Amazon, and Telesat are building large, low earth orbit (LEO) satellite networks that provide global Internet access without the need for terrestrial fiber.
The Starlink constellation is already partly deployed in a test phase~\cite{Pultarova2015-ml}.

\begin{figure}
    \centering
    \captionsetup[subfigure]{justification=centering}
    \begin{subfigure}{.5\columnwidth}
        \centering
        \includegraphics[height=3.5cm]{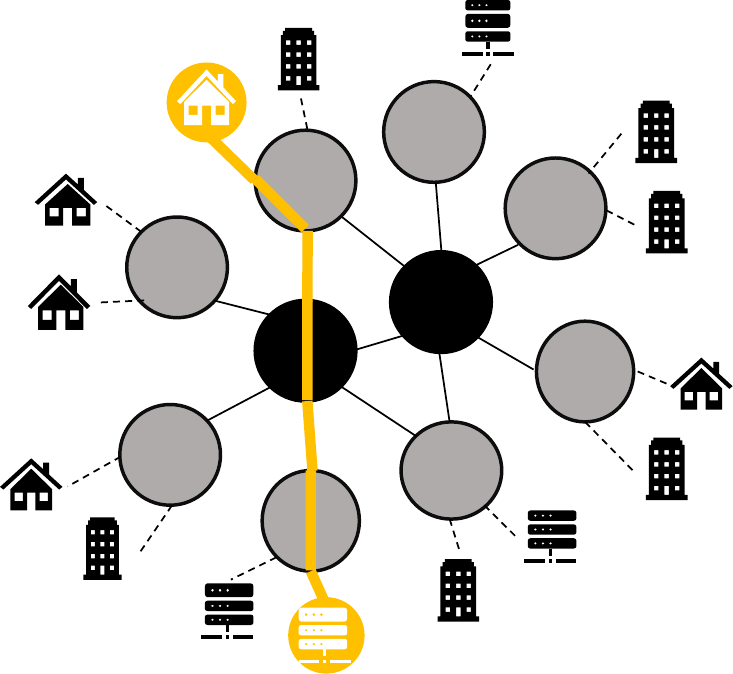}
        \caption{Hierarchical, tiered topology}
        \label{fig:topology_hierarchical}
    \end{subfigure}%
    \begin{subfigure}{.5\columnwidth}
        \centering
        \includegraphics[height=3.5cm]{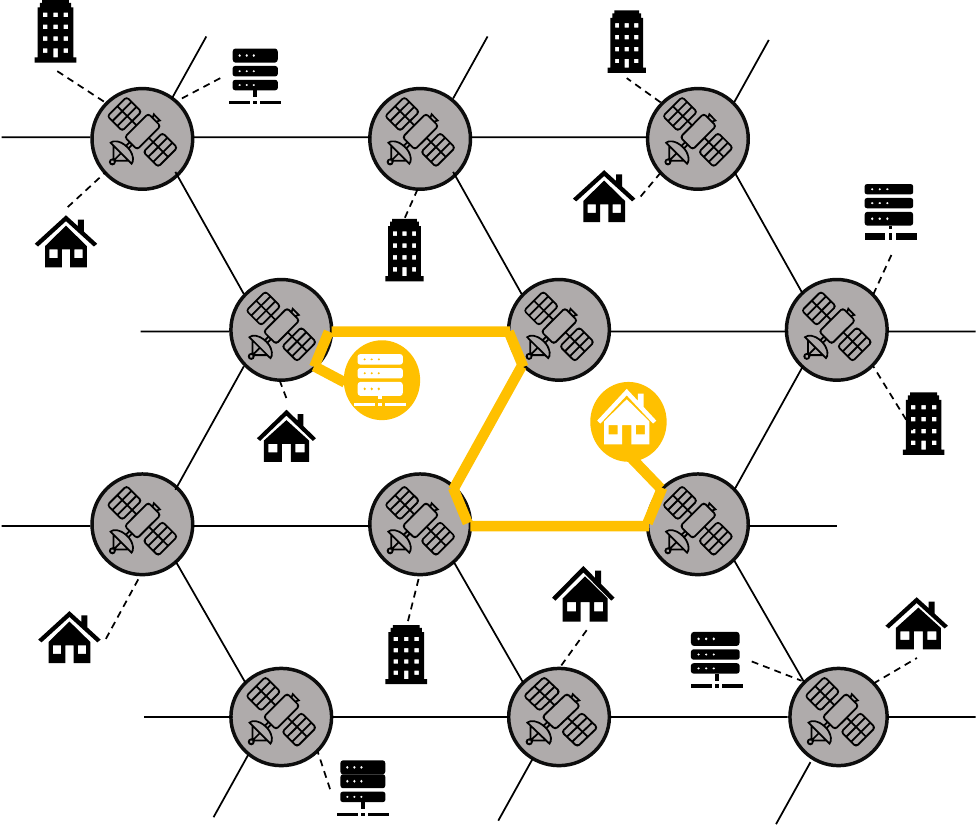}
        \caption{Distributed, dynamic topology}
        \label{fig:topology_sat}
    \end{subfigure}%
    \caption{While requests currently traverse hierarchically arranged tiers of networks (Figure~\ref{fig:topology_hierarchical}), all global clients will be able to communicate directly in the shared wide-area satellite network (Figure~\ref{fig:topology_sat}).}
    \label{fig:topology}
    \vspace{-0.25cm}
\end{figure}

These new satellite connection networks differ tremendously from the traditional tiered topology of the Internet, as we illustrate in Figure~\ref{fig:topology}.
Currently, users connect to the Internet via Tier 3 networks provided by their local ISPs, which connect to Tier 2 networks and the backbone of only a handful of Tier 1 networks.
Operators of content delivery networks (CDNs) exploit this hierarchical topology: by placing their points-of-presence (PoP) within Tier 3 networks that group clients in the same vicinity, they can serve many clients with low access latency by replicating web content close to its consumers~\cite{AkamaiTechnologiesundated-ro,Sivasubramanian2004-eo}.

In satellite access networks, all consumers have direct or near-direct access to the global satellite backhaul network.
Additionally, as the individual satellites are not geostationary, they continuously connect to different ground stations as they orbit over the earth.
The distributed and dynamic network presents a significant challenge for CDN operators as their placement of PoPs can no longer exploit a hierarchical network topology.
While CDNs for satellite networks have been proposed as potential use-cases for orbital edge computing~\cite{Bhosale2020-aa,Bhattacherjee2020-kr}, to the best of our knowledge, PoP placement has so far not been investigated.

In this paper, we propose four novel approaches to PoP placement in large LEO satellite constellations and evaluate them through simulation.
To this end, we make the following core contributions:

\begin{itemize}
    \item We propose four different PoP selection strategies for satellite access networks (Section~\ref{sec:strategies}).
    \item We present a simulation environment for web requests in satellite networks and use it to evaluate our four strategies (Section~\ref{sec:evaluation}).
    \item We discuss their implications for future development of large LEO satellite networks (Section~\ref{sec:discussion}).
\end{itemize}

%% file: Sections/2_background.tex
\section{Background}
\label{sec:background}

In this section, we briefly introduce and describe the state of the art in large LEO satellite communication networks and CDNs.
The remainder of this paper is based on this terminology.

\subsection{Large LEO Satellite Communication Networks}

While satellite-backed Internet access using geostationary satellites at an altitude of 35,000km has been in operation for decades, the induced communication latency makes it infeasible for most use cases~\cite{Clarke1945-qb,Iida2000-il}.
Currently, however, SpaceX, Amazon, and others are designing new, large LEO satellite communication networks that comprise thousands of satellites that orbit the earth at a much lower altitude of less than 600km to provide global Internet access~\cite{Pultarova2015-ml}.
Rather than only relaying signals between two ground stations, these satellites feature inter-satellite links (ISL) that facilitate network connection between satellites.
Due to the surrounding vacuum, ISLs can leverage a 50\% faster speed of light compared to fiber cables, and the path from ingress to egress satellite is more direct than traversing the traditional tiers of the global Internet.
Consequently, two ground stations can not only communicate directly through this satellite network but can also do so with reduced latency compared to terrestrial fiber~\cite{Khan2015-wf}.
Especially SpaceX market this new ``space Internet'' not just as an option for locations without access to terrestrial fiber but as an alternative that outperforms conventional connection methods~\cite{Del_Portillo2019-al,Giuliari2020-pj,Klenze2018-og,Bhattacherjee2018-vc,Bhattacherjee2019-jz}.

\begin{figure}
    \centering
    \includegraphics[width=\linewidth]{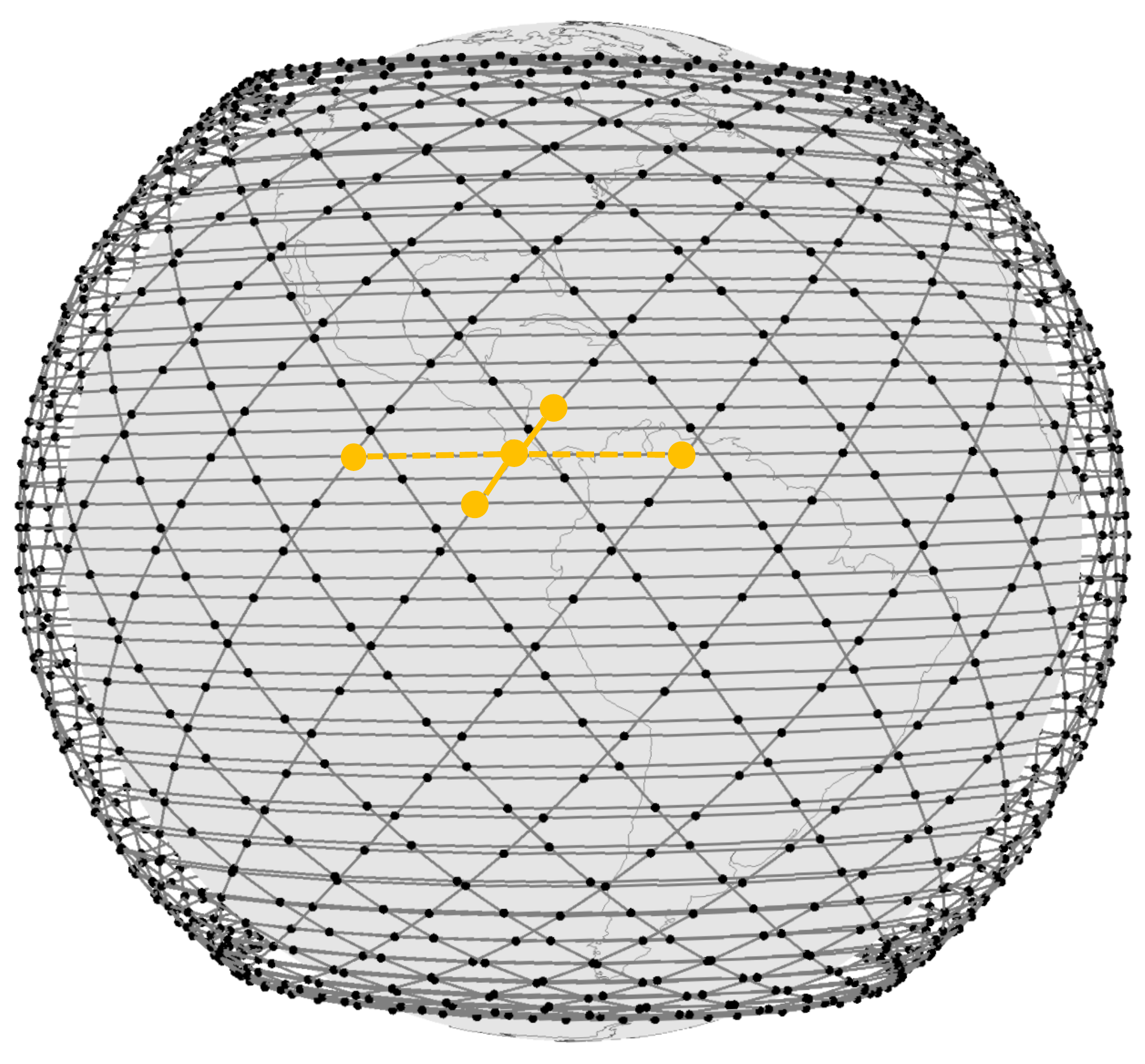}
    \caption{Phase~\textrm{I} of the planned Starlink constellation developed by SpaceX with 24 planes of 66 satellites each, orbit inclination of 53°, and altitude of 550km. To show the +Grid ISLs we mark the links for one satellite as an example~\cite{Bhattacherjee2019-jz,Mark_Handley2018-de}.}
    \label{fig:constellation}
    \vspace{-0.25cm}
\end{figure}

In these large LEO satellite constellations, satellites are arranged in planes evenly spaced around the earth, with each plane comprising an evenly-spaced group of satellites in the same orbit.
To give an example, phase \textrm{I} of the planned SpaceX constellation comprises a total of 1,584 satellites, with 24 planes of 66 satellites each (Figure~\ref{fig:constellation}).
The orbits each have an inclination of 53° and an altitude of 550km~\cite{Bhattacherjee2019-jz,Mark_Handley2018-de}.
An orbit's inclination describes the angle of its plane compared to the earth's equatorial plane, while the orbit's altitude is its distance to the surface of the earth~\cite{nasa-dq}.
As the constellation uses only circular orbits, a satellite's altitude remains constant over its orbital period.

Here, the ISLs are likely to be arranged in a neighbor-grid, or \emph{+Grid} pattern, where each satellite keeps links to its successor and predecessor in its plane in addition to two cross-plane links to neighboring satellites in both adjacent planes.
The constellation imposes a connected grid over the globe, covering each ground point with access to the network and enabling routing between any two ground terminals over the satellite network~\cite{Bhattacherjee2019-jz,Klenze2018-og,Mark_Handley2018-de,Handley2019-ce}.

A data item sent from a server to a client over the satellite network is first passed from the server to a satellite uplink.
Data centers are likely to be equipped with one or more of these uplink dishes to provide fast and direct access to the satellite network to serve clients better.
Second, the data item is sent to the optimal satellite.
This might be the nearest satellite with the best visibility or connection, or it could be the one that is the closest to the target location of the request to optimize latency.
It passes through the satellite network over ISLs to the satellite with a connection to the target ground station.
Finally, the item is passed from the satellite network to a satellite dish on the ground, from where it reaches the target computer.

An as-of-yet unknown aspect is how these ground stations will be deployed: while a Starlink satellite dish, for example, is too big for typical consumer hardware such as a mobile phone, it could be installed on private houses, covering only a few tens of devices within that house; as an uplink for a radio tower providing network access for hundreds of devices at a time; or even on a city or communal level for thousands of connected devices.
Even a hybrid of these installation options is possible, depending on the needs and available resources in different areas.
The tradeoff between managing only a few high-bandwidth connections or lots of low-bandwidth connections per satellite is likely to be handled differently on a provider-by-provider basis~\cite{Handley2018-ay,Bhattacherjee2018-vc,Bhattacherjee2019-jz}.

\subsection{Replicating Web Content in Content Delivery Networks}

CDNs replicate and distribute web documents of different web sites at well-chosen locations close to clients and redirect client requests to these locations.
These locations at the edge of the Internet are referred to as points-of-presence (PoPs) of the CDN~\cite{Tanenbaum2016-jp,Sivasubramanian2004-eo}.

\begin{figure}
    \includegraphics[width=\linewidth]{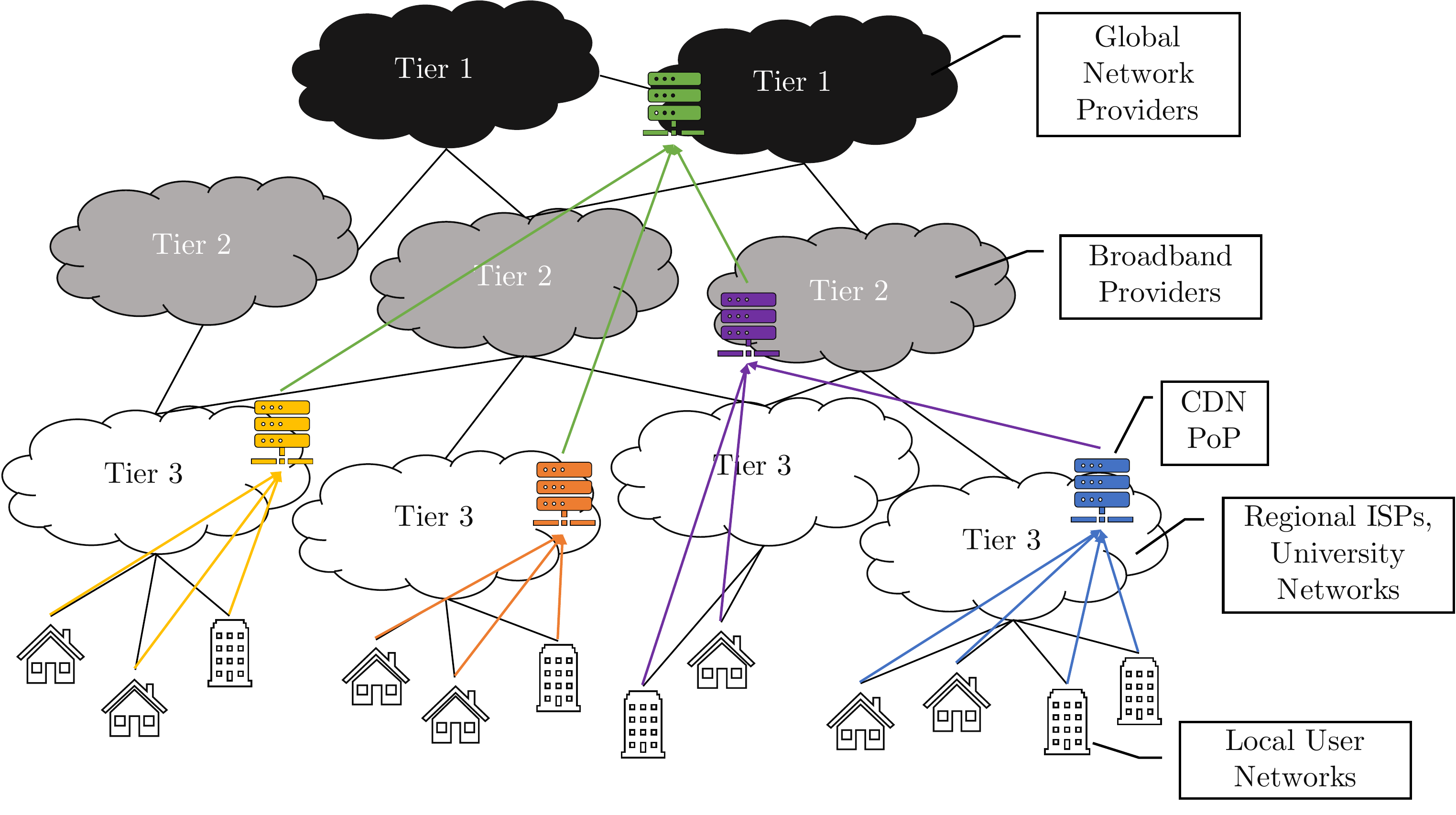}
    \caption{CDN PoPs in the tiered Internet. While most clients are served from PoPs within their ISP access networks, additional PoPs serve networks where no capacity has been allocated and are used for tiered distribution of content~\cite{AkamaiTechnologiesundated-ro}.}
    \label{fig:cdnpops}
    \vspace{-0.25cm}
\end{figure}

Placing these PoPs close to clients reduces access latency and minimizes the required bandwidth in the network.
Simultaneously, to save on operational costs, the CDN operator benefits from operating as few PoPs as possible, i.e., PoPs should also cover many potential clients at the same time.
As a result, PoPs are often placed in Tier 3 networks that directly serve end-users.
We show an example of PoP placement in the tiered Internet infrastructure in Figure~\ref{fig:cdnpops}.
For fault tolerance and tiered distribution, additional PoPs can also be installed in Tier 2 or Tier 1 networks~\cite{AkamaiTechnologiesundated-ro}.
For example, Akamai, one of the largest CDN operators, operates approximately 300,000 servers in more than 130 countries~\cite{akamai-rn}.

Placing compute and storage PoPs at the edge of the networks is also the basis of today's edge and fog computing~\cite{paper_bermbach_fog_vision}.

%% file: Sections/3_pop_strategies.tex
\section{Strategies for PoP Selection}
\label{sec:strategies}

When deploying a CDN for LEO satellite-based Internet, significant savings in terms of bandwidth usage and improved access latency for clients can be achieved.
The critical question is where to put the CDN's data, i.e., to choose the PoPs.
In this section, we introduce four novel strategies for selecting such CDN PoPs in networks that rely on the novel large LEO satellite constellations: ground station PoPs (\emph{GST}), simple satellite PoPs (\emph{SAT}), satellite PoPs with time-to-live (\emph{SAT-TTL}), and satellite PoPs with internal replication (\emph{SAT-REP}).

\subsection{Ground Station PoPs (GST)}

As ground station hardware is too large for end-user devices, a ground station will usually act as a gateway for several devices.
Given that all devices served by a single ground station would be located in the same general area, they are likely to also access similar content~\cite{Hasenburg2020-xi,Hasenburg2020-gf,DOro2014-dk,paper_hasenburg_geobroker}.
These two factors make ground stations suitable PoPs when serving web content via the satellite network.
Furthermore, deploying hardware, i.e., storage devices, on the ground is no challenge and comparatively cheap.

In GST, content is requested from the local ground station and, if available, served from there.
If the content is not already stored locally, it is fetched from the origin location over the satellite network, and a copy is stored in the ground station's local store.
This PoP placement strategy decreases bandwidth usage and is most efficient when there is a high ratio of devices to ground stations.

\subsection{Simple Satellite PoPs (SAT)}
\label{subsec:localsatellite}

With many ground stations deployed and few devices per ground station, e.g., with every household using their own satellite dish to connect to the network, placing PoPs in the ground stations may not offer significant advantages over on-device caching.
In such scenarios, request paths from multiple ground stations, and thus from multiple end-user devices, intersect on a satellite, so that we must consider satellites as possible PoPs for content replication.
Of course, deploying storage hardware within these satellites is a considerable challenge compared to deploying that same hardware at ground stations.
We further discuss this challenge in Section~\ref{sec:discussion}.

The naive approach to satellite PoPs keeps a local copy of data items on the first satellite that the requesting ground station connects to.
Subsequent requests can then be served from this PoP.
This does not make use of any ISLs and has no negative impact on ISL bandwidth consumption.

The main challenge of this approach is that satellites are not geostationary:
A satellite storing a local copy of a data item requested on one side of the planet will move to the opposite side of the earth within an hour, and the data item may be of no use there.
If the item is popular enough that it is requested several times within a few minutes, or if the requested item is relevant to large geographic areas, this could still provide a sufficient increase in efficiency.
Moreover, when the satellite has orbited the planet once, its locally stored data becomes relevant again.

\subsection{Satellite PoPs with Time-to-Live (SAT-TTL)}

\begin{figure*}
    \begin{subfigure}{.33\textwidth}
        \centering
        \includegraphics[width=.9\linewidth]{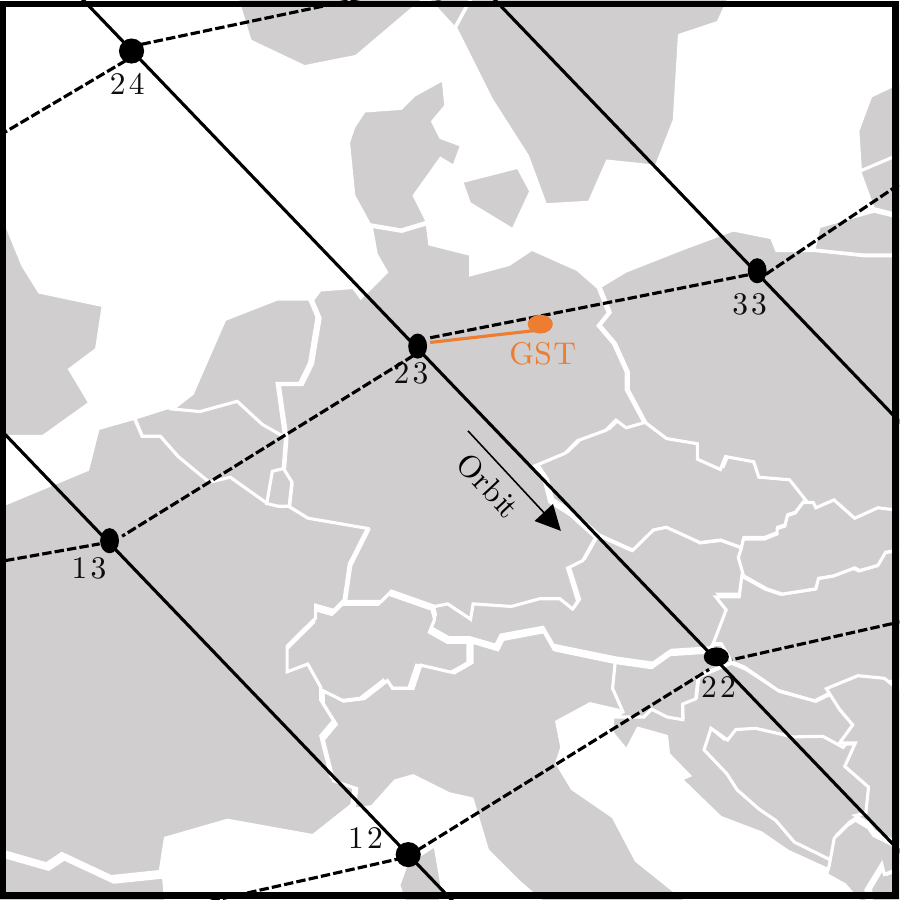}
        \caption{initial constellation}
        \label{fig:flying1}
    \end{subfigure}%
    \begin{subfigure}{.33\textwidth}
        \centering
        \includegraphics[width=.9\linewidth]{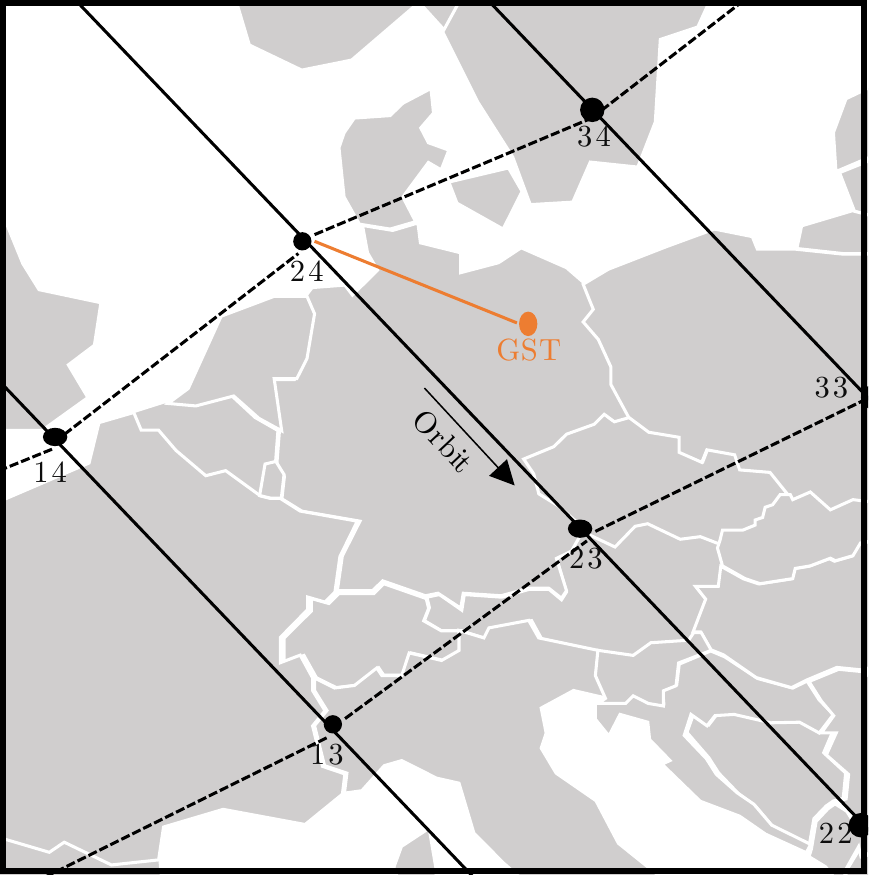}
        \caption{handoff within a plane}
        \label{fig:flying2}
    \end{subfigure}%
    \begin{subfigure}{.33\textwidth}
        \centering
        \includegraphics[width=.9\linewidth]{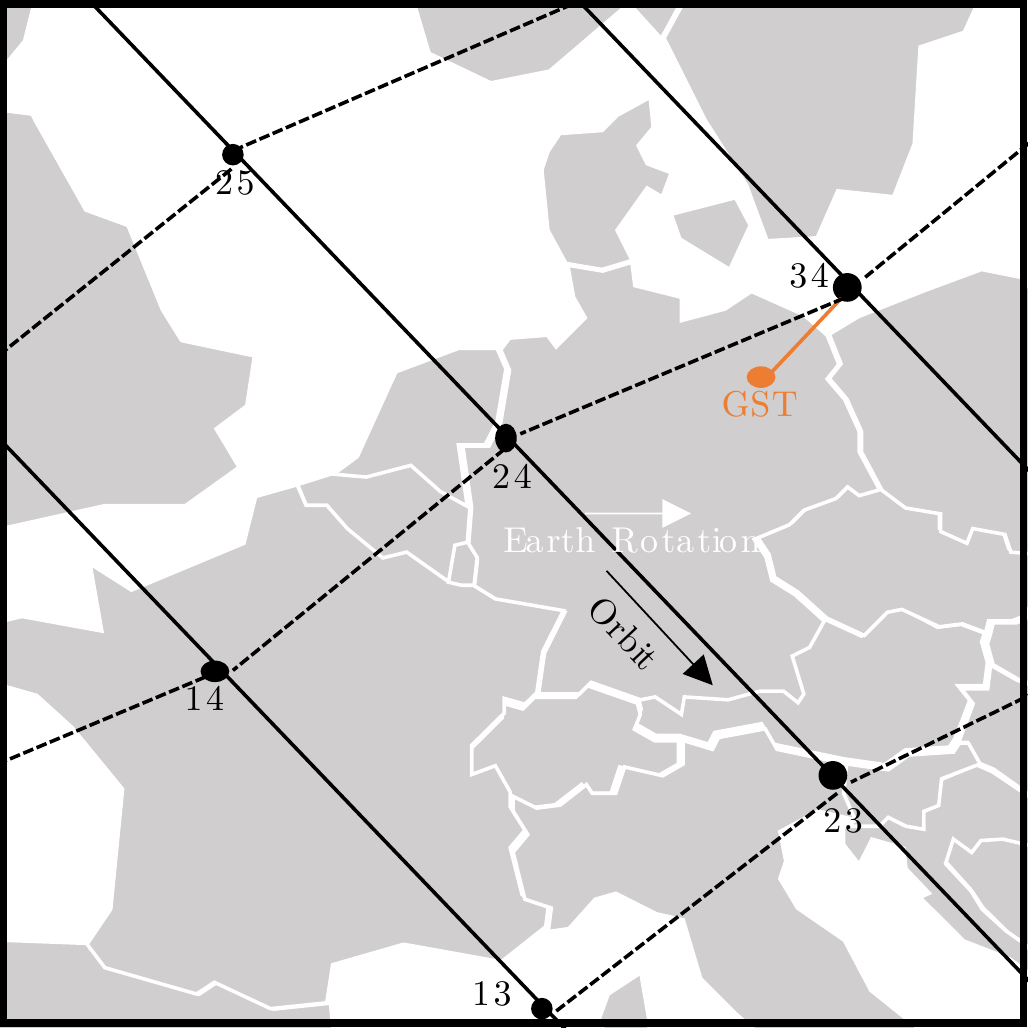}
        \caption{handoff across planes}
        \label{fig:flying3}
    \end{subfigure}
    \caption{Handoff between satellites as the satellites orbit around the earth and the earth rotates underneath the constellation. Once a ground station loses the connection with a satellite, it connects to that satellite's successor in the same plane or to a neighbor in the adjacent plane.}
    \label{fig:flying}
    \vspace{-0.25cm}
\end{figure*}

To forego the issue of a satellite ``carrying'' a locally replicated data item to another part of the world where different content is required and thereby unnecessarily blocking its limited storage, we introduce a time-to-live (TTL).
This TTL does not apply to individual data items but rather to the complete local PoP storage.

The local store of a satellite PoP should be purged as soon as a ground station that it serves connects to a new satellite so that only replicas for a single location are kept on a satellite.
We show this handoff in Figures~\ref{fig:flying1} and~\ref{fig:flying2}.
As the satellites are evenly spaced within a plane, this handoff happens at a constant interval -- the satellites' orbital period divided by the number of satellites within one plane.
Additionally, a ground station connects to a new plane every once in a while as the earth rotates underneath the satellite constellation (Figure~\ref{fig:flying3}).
This duration is derived from the time one full rotation of earth takes, i.e., one day or 86,400s, and the number of planes in the constellation.
However, a full orbit of a LEO satellite is faster, in the order of one to two hours.
Consequently, if the number of satellites per plane is on the same order of magnitude as the number of planes in the constellation (or even smaller as is the case with the phase~\textrm{I} Starlink constellation), it is sufficient to consider the in-plane handoff for the TTL.

\begin{equation}
    \label{eq:ttl}
    T_{TTL} = \frac{T_{orbit}}{\# satellites / plane}
\end{equation}

We thus determine this $T_{TTL}$ as shown in Equation~\ref{eq:ttl}.
In the case of Starlink with an orbital period of 5,730s and 66 satellites within every plane, $T_{TTL}$ would be 86.8s.
To compare, a cross-plane handoff with the 24 planes in the constellation happens only once every 3,600s.

Compared to the SAT strategy, this PoP strategy leads to less storage required at each satellite, yet also requires more bandwidth.
After each duration of $T_{TTL}$, if a data item is requested, it has to be fetched from the origin servers to keep replicas ready for subsequent requests.
Assuming a strictly location-based demand of data items, the outcome of SAT-TTL should be comparable to LRU caching, where clients frequently connect to a new cache.

\subsection{Satellite PoPs with Internal Replication (SAT-REP)}

Fetching data from the origin server after each TTL expiration of SAT-TTL may, however, lead to unnecessary bandwidth usage.
Alternatively, it may be possible to use the ISLs in the satellite constellation:

As satellites within a plane and the planes themselves are evenly spaced within the constellation, it is always possible to uniquely identify the next satellite a ground station connects to and to preemptively propagate the local PoP store to that satellite.
As every satellite has a direct ISL to its successor and predecessor within its plane, this propagation requires only a single hop, whereas fetching data items from the origin server again could require multiple hops.

\begin{figure*}
    \begin{subfigure}{.33\textwidth}
        \centering
        \includegraphics[width=.9\linewidth]{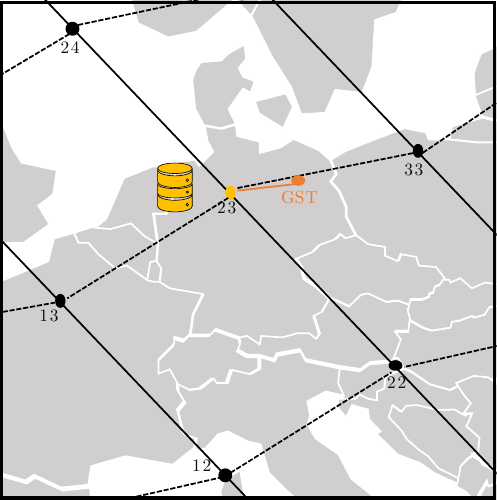}
        \caption{$T_{0}$}
        \label{fig:crossreplica1}
    \end{subfigure}%
    \begin{subfigure}{.33\textwidth}
        \centering
        \includegraphics[width=.9\linewidth]{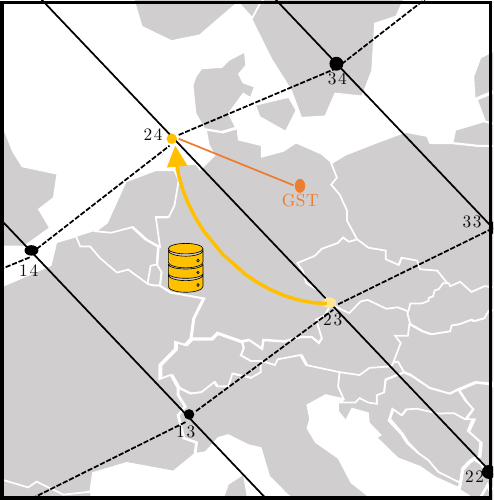}
        \caption{$T_{intra}$}
        \label{fig:crossreplica2}
    \end{subfigure}%
    \begin{subfigure}{.33\textwidth}
        \centering
        \includegraphics[width=.9\linewidth]{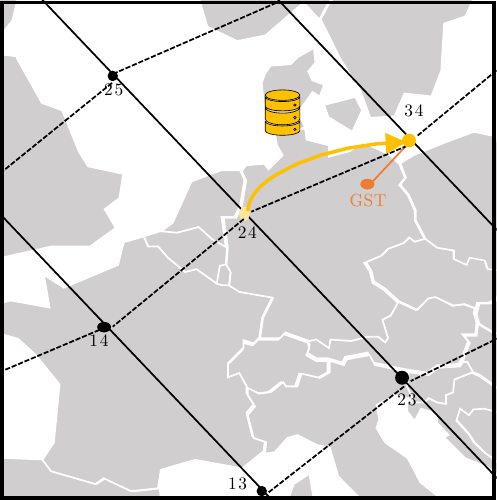}
        \caption{$T_{cross}$}
        \label{fig:crossreplica3}
    \end{subfigure}
    \caption{Replication of data in the SAT-REP strategy as the satellites orbit around the earth and the earth rotates underneath the constellation. When a ground station is connected to a satellite at $T_{0}$ and serves requests of that ground station from its local replica, it propagates its local store to either its successor in the same plane at $T_{intra}$ or to its neighbor in the adjacent plane at $T_{cross}$ depending on earth rotation.}
    \label{fig:crossreplica}
    \vspace{-0.25cm}
\end{figure*}

In this strategy, each satellite PoP removes the locally stored data items after the TTL expires yet first propagates its entire local store to its next satellite.
We illustrate this for handoffs within a plane in Figures~\ref{fig:crossreplica1} and~\ref{fig:crossreplica2}.
Hence, a ground station is always connected to a satellite PoP with local copies of the data items it requires without fetching that data from the origin location again.

The additional challenge here, however, is the rotation of the earth underneath the satellite location.
Preemptively propagating data items within a single plane works only as long as that plane is the closest to the ground station.
As the earth rotates, the ground station slowly moves towards the next plane of satellites.
Consequently, an additional cross-plane propagation has to be initiated periodically, which we show in Figure~\ref{fig:crossreplica3}.
While we argue that this is negligible for the SAT-TTL strategy, it does make a difference here as the local replica set is long-lived.

\begin{subequations}
    \begin{equation}
        \label{eq:propagation_intra}
        T_{intra} = \frac{T_{orbit}}{\# satellites / plane}
    \end{equation}
    \begin{equation}
        \label{eq:propagation_cross}
        T_{cross} =  \frac{86400s}{\#planes}
    \end{equation}
\end{subequations}

We thus consider two points in time for propagation of the local PoP store to a different satellite: $T_{intra}$ for intra-plane propagation and $T_{cross}$ for cross-plane propagation, as shown in Equations~\ref{eq:propagation_intra} and~\ref{eq:propagation_cross}, respectively.
$T_{intra}$ is calculated in the same way as the $T_{TTL}$ in the SAT-TTL strategy, by dividing the orbital period among the evenly spaced satellites of a plane.
Cross-plane propagation at $T_{cross}$ is the duration of one full rotation of the earth, 24 hours, divided by the number of planes in the constellation.
For example, in the phase~\textrm{I} Starlink constellation with 24 evenly spaced planes, such a cross-plane replication would occur every 3,600s, or every hour, while intra-plane propagation would occur every 86.8s.

Compared to the SAT strategy, this approach uses less storage as replicas of data items are only stored in satellites connected to the ground stations that request these items.
Proactively propagating local storage also means that fewer items have to be fetched from the origin servers compared to both the SAT-TTL and the SAT strategy.
However, the frequent propagation requires more bandwidth than in the SAT and SAT-TTL strategy, where every satellite's local replica store acts independently.
In this way, every strategy is a tradeoff decision, which we explore through simulation in the next section.

%% file: Sections/4_simulation.tex
\section{Evaluation}
\label{sec:evaluation}

To evaluate our four proposed CDN PoP selection strategies in addition to the default strategy where only the origin server keeps content replicas, we simulate a full day of web requests in a large LEO satellite network.
We shortly describe our simulation environment, introduce our simulation tool, and present the results.

\subsection{Simulation Scenario}

We run two experiments in which we simulate the phase~\textrm{I} Starlink constellation as described in~\cite{Bhattacherjee2019-jz}, which comprises 24 planes of 66 satellites each and provides Internet access to consumers in the form of ground stations on earth.
Over 24 hours, clients request items from an origin location in one-second intervals.
These requests are first sent to their ground station's nearest satellite.
They are then routed to the satellite closest to the origin server's ground station.
This request can be intercepted by a ground station or satellite with a local copy of that data item, depending on the PoP selection strategy which we evaluate.
We show an overview of the parameters of our two experiments in Table~\ref{tab:parameters}.

\input{Tables/parameters}

Please, note that our work aims to evaluate PoP selection strategies; hence, we need to abstract from the orthogonal problem of cache replacement algorithms in local stores of the PoPs.
We thus assume in our simulation experiments that a local store can grow infinitely at every PoP.
While this is unrealistic in real deployments, it helps us understand the amount of different data items a node has to handle without simulating different store sizes and cache replacement algorithms.
It can be argued that choosing the correct store size and replacement algorithm for a PoP depends on the amount of data items a PoP has to manage, i.e., it is a result of our work.

We run our simulation on a country-level, i.e., in every simulation we use a list of cities and their respective population of a single country.
The assumption is that the requested items are similar within a single country~\cite{DOro2014-dk,Hasenburg2020-xi,Hasenburg2020-gf}.
To compare the impact of country size, we use data sets of two countries that we simulate separately.
We use a dataset of cities in the US with a population larger than 40,000, generated from the R \texttt{maps} data set\footnote{\url{https://github.com/adeckmyn/maps}}.
Here, cities are spread over a large landmass with most of the population, i.e., clients on either coast of the country.
For comparison, we use a set of all cities and towns in Switzerland, which is a significantly smaller country.
This dataset is based on \emph{OpenStreetMaps} data\footnote{\url{https://openstreetmaps.org}}.
As our data origin we choose a single ground station within the country we simulate.
When simulating our proposed strategies, we assume a tiered content delivery; the individual PoPs, for example the satellites in the SAT strategy, pull data from this single ground station, save a copy locally, and later serve this copy.

We also consider different client numbers for ground stations as this influences the GST strategy's effectiveness:
First, we assume 10,000 clients per ground terminal, i.e., a neighborhood or university campus sharing a ground station.
Second, we run the simulation with only 100 clients per terminal, which corresponds to a cell tower or larger building.
Third, we assume ten clients sharing one ground station, such as a single household or small business.
We show individual results as GST-10000, GST-100, and GST-10, respectively.

\subsection{Simulation Tool}

We extend the \emph{SILLEO-SCNS} routing simulator presented in~\cite{Kempton2020-qx}, which was written in Python3 and uses the \texttt{PyAstronomy} and \texttt{python-igraph} packages.
Our improved and extended version is available as open-source\footnote{\url{https://github.com/pfandzelter/LLEOSCN-CDN-Sim}}.

First, we added a workload generator that uses the distribution functions for requests and data in a small item cache in a CDN as identified by~\cite{Shafiq2016-kj}.
The analyzed CDN serves terrestrial users, yet we have no reason to believe that satellite Internet is used differently than its terrestrial counterpart.
The workload generator creates a set of data items in a specified size range and with specific popularities, and then generates requests from given client locations.

Second, our simulation tool produces a list of traces for each time step, where each trace is a client request with client ground station, ISL path, server ground station, and the item size, which defines the required bandwidth for that request.

Third, we added a separate CDN replication step that takes the request traces and simulates where, how, and when replicas would be stored for the different PoP strategies.
From here, we derive our final results as the storage and bandwidth requirements for the different PoP strategies can be calculated.

\subsection{Results}
\label{sec:results}

We present our results in two parts: the bandwidth usage in the network and storage requirements at the PoPs.

Although both simulation experiments simulate a full day, we present only a small excerpt of 10 minutes in each graph:
After an initial ramp-up period, the patterns seen over the experiment duration are mostly constant, and this smaller view is useful for a more detailed analysis.

\begin{figure*}
    \begin{subfigure}{.5\textwidth}
        \centering
        \includegraphics[width=1\linewidth]{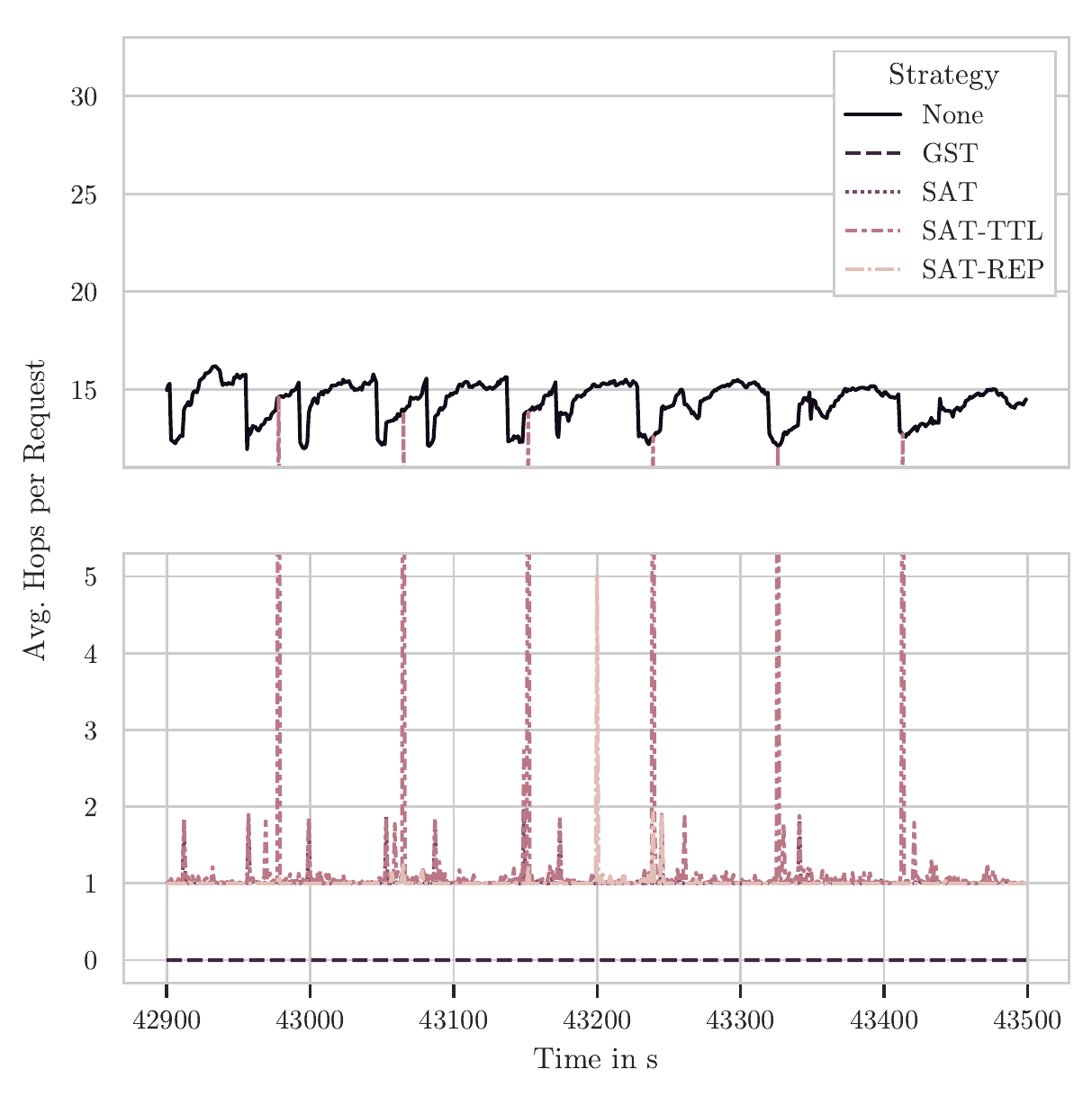}
        \caption{US locations}
        \label{fig:hops_us}
    \end{subfigure}%
    \begin{subfigure}{.5\textwidth}
        \centering
        \includegraphics[width=1\linewidth]{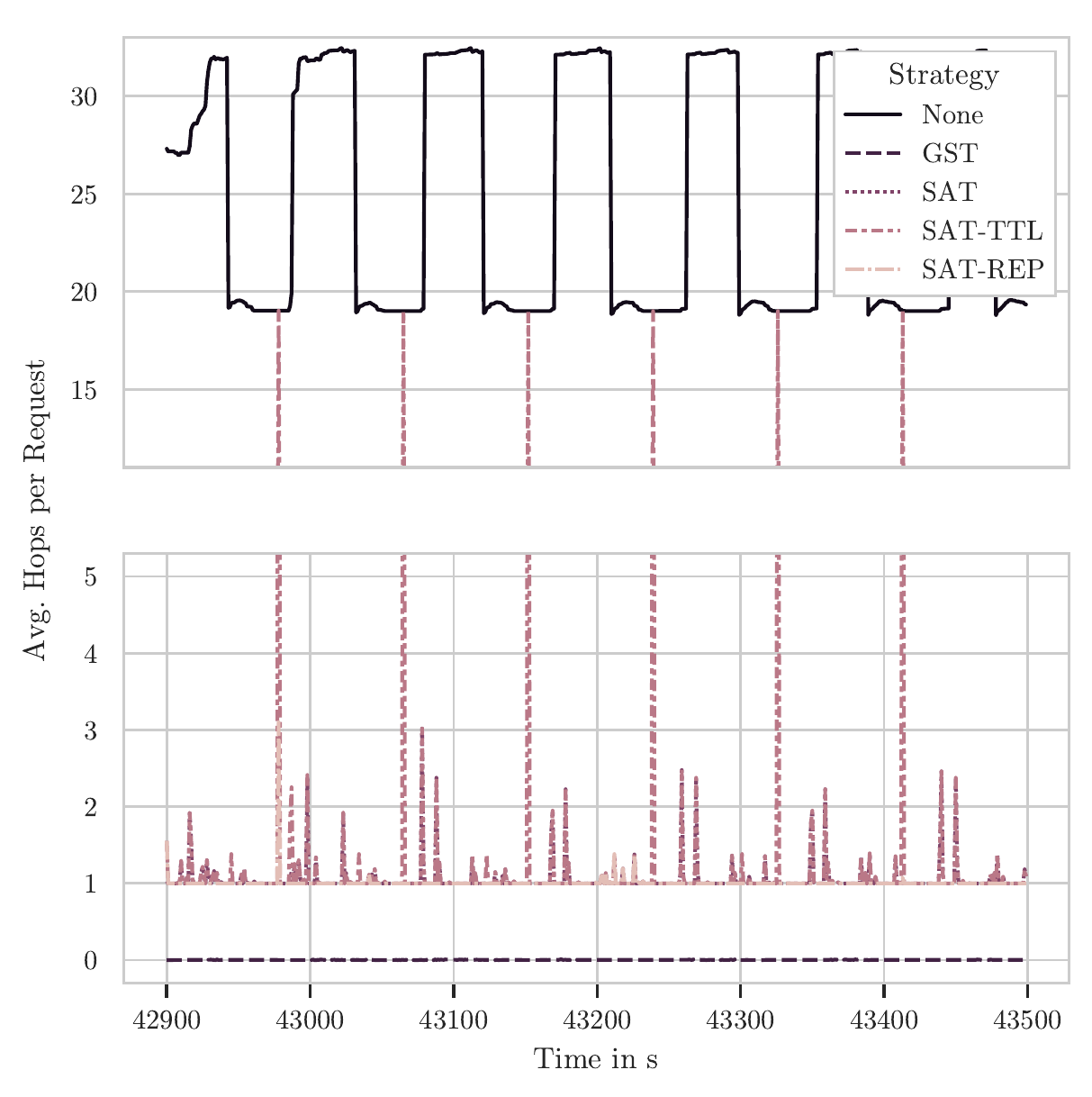}
        \caption{Swiss locations}
        \label{fig:hops_swiss}
    \end{subfigure}
    \caption{Average number of hops per request with different strategies.}
    \label{fig:hops}
    \vspace{-0.25cm}
\end{figure*}

\textbf{Bandwidth:}
The amount of bandwidth used is the main target of optimization through CDN PoPs in our case.
Bandwidth usage caused by a request depends on two factors: the requested item's size and the number of hops that the request needs.
As the size of each item does not change with different PoP selection strategies, we look at the number of hops needed by requests to estimate the strain on the network, which we show in Figure~\ref{fig:hops}.
For hops, we consider only those that pass through the satellite network, i.e., between two ground stations, and disregard additional hops between user devices and ground terminals.
Note that the bandwidth usage is proportional to the end-user latency for accessing data items when we assume comparable network latency for all network hops which is realistic for the ISL of the simulated LEO constellation.

For the simulation without any PoPs within the network, we observe that the average number of hops per request fluctuates between 12 and 16 for the US location set and between 18 and 33 for the Swiss ground station set.
This fluctuation and unexpectedly high hop count, especially for the small Swiss data set, is caused by the complex network topology of the satellite networks, as described by Handley~\cite{Handley2018-ay}.
Over any location covered by the LEO constellation, planes from opposing sides of the planet cross over, with one moving in a North-East direction and the other in a South-East direction.
As a ground station connects to the closest satellite, it is possible that two ground stations that are physically close to one another connect to opposing planes and thus have a considerable network distance.

For the GST strategy, we observe that 99\% of all requests are served from the ground station's local store after a short period.
Here, the average hop count is close to zero after the initial ramp-up period.
After a few requests to the most popular items have been made, less than 1\% of items have to be fetched from the origin location and complete the full path through the network.
Given the items' popularity, this happens quickly after the start of the simulation.
At this point, a majority of requests can be served from the ground station PoPs.
To give an example from the simulation experiment with locations in Switzerland, 99\% of the requests are consistently served from PoPs after a simulated duration of 4s, 274s, and 2,680s with 10,000, 100, and 10 clients per ground station, respectively.
While this is a positive effect for hop counts, it has adverse effects for the local store at the ground station PoPs, as we show in the next part.

For the SAT, SAT-TTL, and SAT-REP strategies, we also see high PoP hit ratios between 95\% and 100\%, yet as each request has to be sent to the satellite first, the average item request takes one hop.
The exception is the SAT-TTL at the expiration of the satellite local store's TTL every 87s.
Here, as the local store on every satellite is emptied, the full path from client ground stations to origin server ground stations has to be traversed to fetch the requested data items.
After these items have been fetched and replicas are stored in the new satellites' stores, requests can again be served from the satellite PoPs.

In the SAT-REP strategy, additional bandwidth is required to propagate the local satellite stores to other satellites.
As local satellite stores are small, however, we find this impact to be negligible compared to the constant bandwidth used by the links between satellites and ground stations.

\textbf{Storage:}
Storing more data items leads to higher requirements for the individual replica servers, which results in a higher investment for the CDN operator.
As such, storage requirements at every CDN PoP should be as low as possible.
Furthermore, in a real deployment, storage would of course be limited -- storage requirements that exceed this capacity limit cannot be fulfilled so that the corresponding data would need to be requested from the origin server.
Thus, the storage requirements are a proxy for how often end-users will benefit from the existence of the CDN.
Again, please, note that we do not limit storage in our experiments which allows us to separate the evaluation of PoP selection strategies from cache eviction policies.

\begin{figure*}
    \begin{subfigure}{.5\textwidth}
        \centering
        \includegraphics[width=1\linewidth]{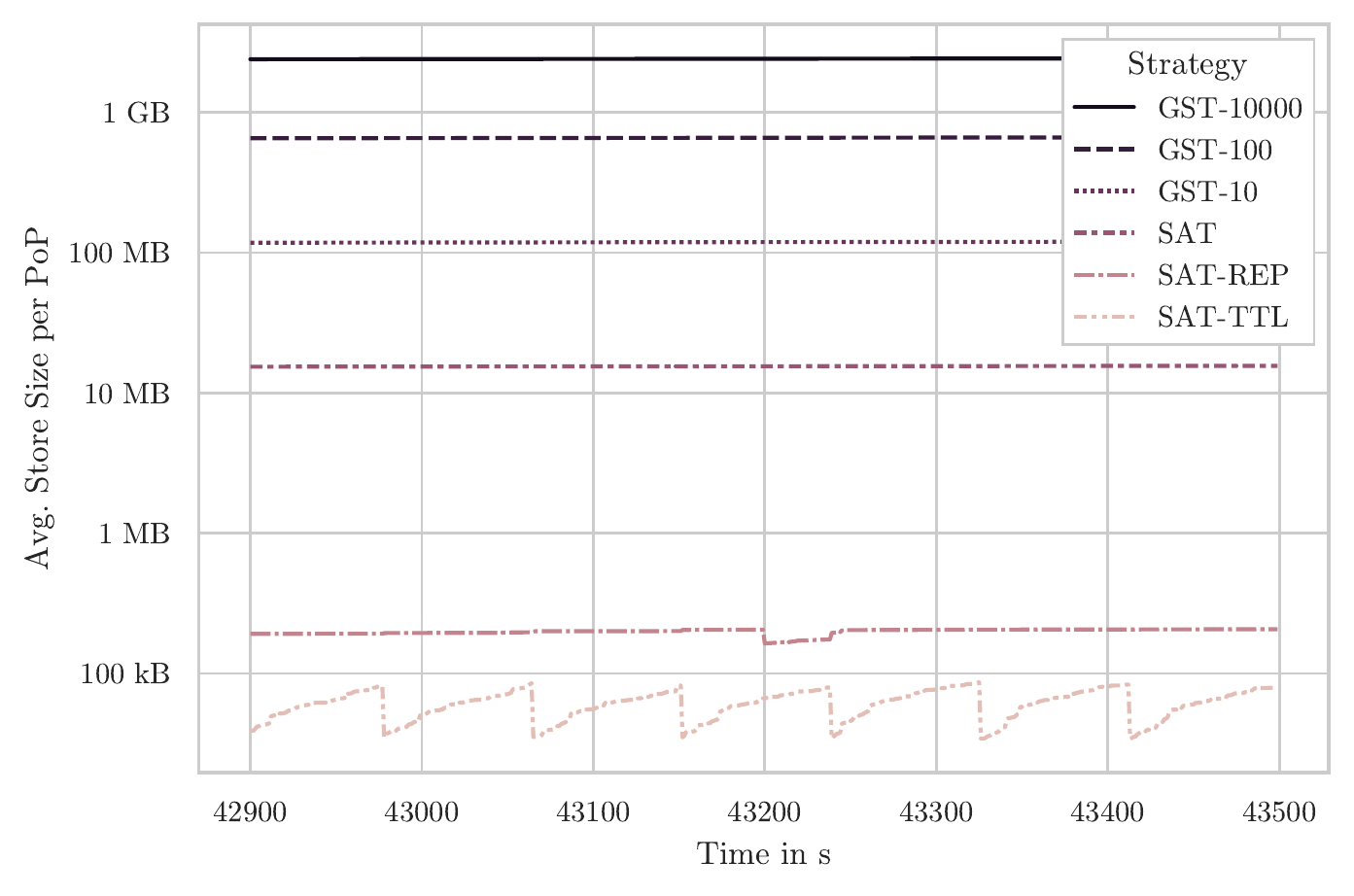}
        \caption{US locations}
        \label{fig:store_us}
    \end{subfigure}%
    \begin{subfigure}{.5\textwidth}
        \centering
        \includegraphics[width=1\linewidth]{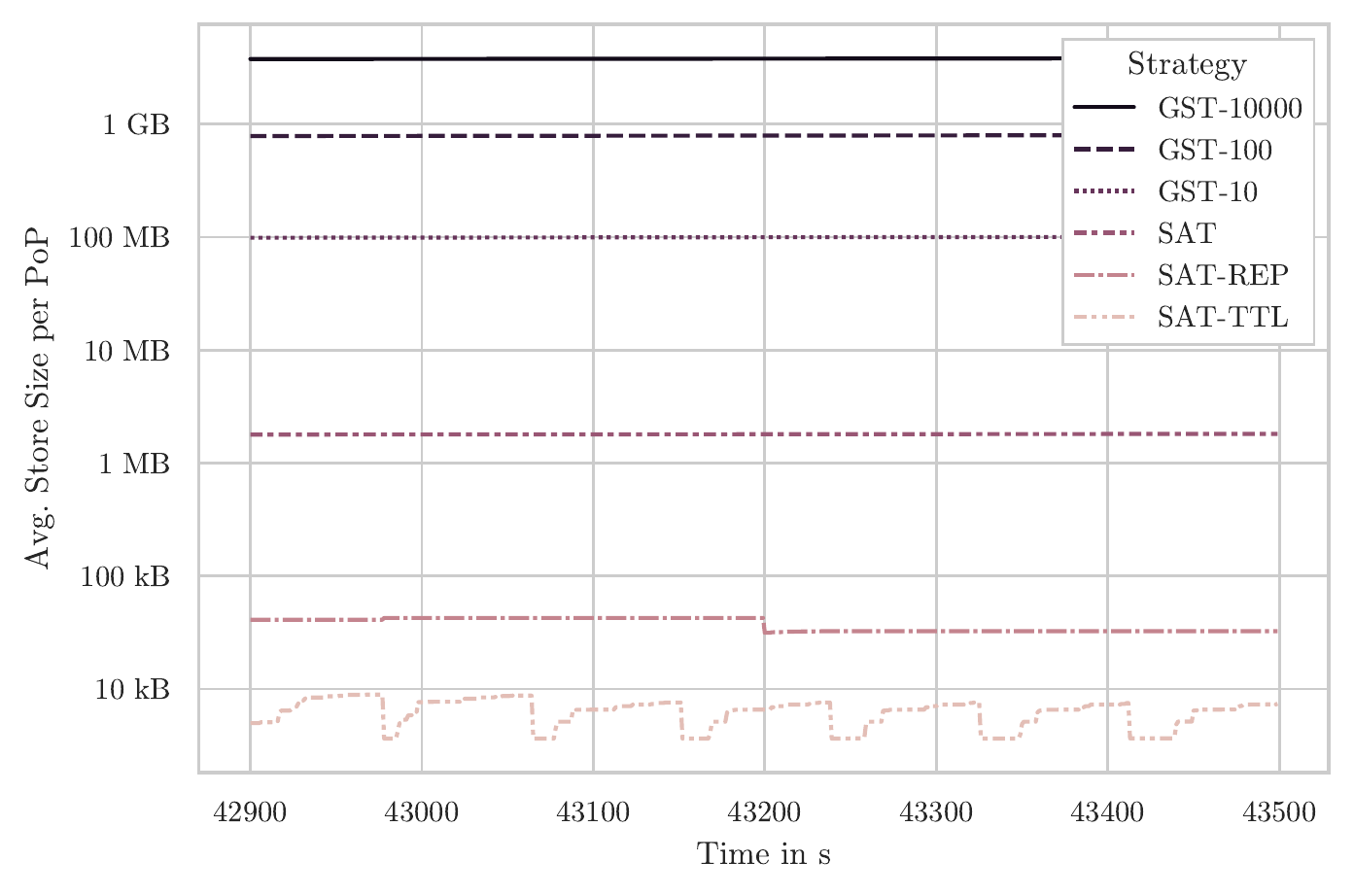}
        \caption{Swiss locations}
        \label{fig:store_swiss}
    \end{subfigure}%
    \caption{Storage use averaged over all PoPs in the constellation with different strategies.}
    \label{fig:store}
    \vspace{-0.25cm}
\end{figure*}

The average storage used across all PoPs in the network is shown in Figure~\ref{fig:store}.
For our baseline test without any PoPs within the satellite network, no storage is necessary, as no data is replicated within the network.

For the GST strategy, we observe that the storage used per node is highly dependent on the granularity of ground stations:
The more clients share a ground station (and, consequently, the fewer ground stations exist in total), the higher the local replicas' average size.
We can explain this effect with the distribution of requests.
The full set of requests is distributed equally across all ground stations.
When the number of ground stations increases, fewer requests are made from each station, which leads to a lower impact of infrequently requested items.
In other words, the more clients a ground station serves, the higher the amount of requests from that ground station, and, subsequently, the higher the total set of unique items present in these requests.

Another intriguing effect here is that the average storage amount per ground station shows no change after the initial ramp-up period as almost all content is served from the ground station PoPs at this time, and only a small percentage of new data items is requested and added to the local stores.
This results in an average storage amount per node that is orders of magnitude higher than with the SAT, SAT-TTL, and SAT-REP strategies.
Additionally, the GST strategy also leads to a significantly higher number of PoPs with this higher average storage requirement (millions of ground stations compared to at most 1,584 satellites), so the total amount of used storage in our CDN is higher as well.

In the SAT strategy simulation, while the average size of the local store within a satellite PoP does not change noticeably, the number of satellites that store information increases continuously.
After simulated 49,675s, no satellites with empty local stores remain in the constellation in the simulation with US locations.
For the smaller set of locations in Switzerland, this duration is higher at 85,117s as ground stations are closer to each other and connect to a smaller total number of satellites at a single point in time.
After this period, all 1,584 satellites have an average store size of 10MB or 1MB in the simulation with US and Swiss locations, respectively.

This fact is important when comparing the SAT-TTL to the SAT-REP (shown in brackets) strategy:
Here, the average storage per node is about 100kB for both strategies in the US simulation and 10kB (100kB) in the Switzerland simulation.
However, the amount of nodes storing data is significantly lower than with the SAT strategy.
On average, for the US simulation only 55 (99) satellites store some data locally.
In the smaller Switzerland simulation, this is the case for only 3 (16) satellites.

\subsection{Implications: Choosing a Strategy}

When selecting PoPs for a CDN network, a tradeoff between optimizing bandwidth usage and allocating storage at PoPs has to be made.
In our case, we have shown two extremes: in our baseline tests without PoPs, no storage is required, yet we see the highest strain on all network links.
In the GST strategy, regardless of ground station granularity, we see that the most popular items are replicated to all ground stations after a short ramp-up period.
This leads to a higher storage requirement but lower bandwidth use, as most requests can be served from the local store.

For the CDN operator, managing this tradeoff comes down to a business decision as both network and storage have costs attached to them.
If network hops are cheaper than storage, sending all content over the network instead of using any PoPs can be the best option.
On the other hand, if storage is cheaper than using bandwidth in the network, reducing hops through many PoPs at ground stations, as with the GST strategy, is better.

While there are many unknowns regarding costs in satellite networks, we expect that a solution somewhere in the middle will be optimal.
Each satellite has a limited routing and bandwidth capacity that cannot be scaled easily, and upgrading a satellite's networking capabilities requires launching a replacement satellite and phasing out the old one, which is costly.
Nevertheless, deploying storage in the network is also a considerable investment.
Low-cost storage hardware in ground stations is feasible, yet the massive scale has to be considered: in our US simulation with 100 clients per ground station, a total of 1.2 million PoPs, one at each ground station, would have to be deployed.
In the simulation, we find that each of those PoPs handles unique content on the order of 1GB, so a considerable chunk of storage would need to be allocated to each PoP.
At a fixed 1,584, the number of satellite-based PoPs is, thus, more manageable.
However, we recognize that deploying storage hardware to space is not a trivial task, given the additional space and power requirements, and the additional maintenance overhead.
Furthermore, in our simulation with the SAT-REP strategy, we find that the average storage requirement per node is significantly smaller than with the GST strategy -- on the order of 100kB.

In practice, we envision the most cost-efficient middle-ground strategy to be a combination of SAT-REP with a reasonable cache eviction policy as in SAT-TTL.
While replication can forego the spikes of bandwidth use after TTL expiration, purging infrequently accessed data items leads to lower storage requirements.

%% file: Tables/parameters.tex
\begin{table}[!t]
    \centering
    \caption{Parameters for the two workloads used in our simulation.}
    \label{tab:parameters}
    \resizebox{\columnwidth}{!}{\begin{tabular}{|l|r|r|}
            \hline
            \textbf{Location Dataset} & \multicolumn{1}{c|}{\textbf{USA}} & \multicolumn{1}{c|}{\textbf{Switzerland}} \\ \hline
            \# Locations              & 996                               & 154                                       \\ \hline
            \# Clients                & 125,736,290                       & 3,246,208                                 \\ \hline
            \# Requests / Second      & 1,000,000                         & 25,000                                    \\ \hline
            Origin Location           & Ashbourne, VA                     & Zurich                                    \\ \hline
            \# Unique Items           & 1,000,000                         & 25,000                                    \\ \hline
            Item Size                 & \multicolumn{2}{r|}{1kB -- 100kB}                                             \\ \hline
            Simulated Time            & \multicolumn{2}{r|}{86400s}                                                   \\ \hline
            Time Interval             & \multicolumn{2}{r|}{1s}                                                       \\ \hline
            \# Planes                 & \multicolumn{2}{r|}{24}                                                       \\ \hline
            \# Satellite / Plane      & \multicolumn{2}{r|}{66}                                                       \\ \hline
            Satellite Altitude        & \multicolumn{2}{r|}{550km}                                                    \\ \hline
            Orbit Inclination         & \multicolumn{2}{r|}{53°}                                                      \\ \hline
        \end{tabular}
    }
    \vspace{-0.55cm}
\end{table}

%% file: Sections/5_discussion.tex
\section{Discussion}
\label{sec:discussion}

Our simulation shows that there are indeed benefits to PoP placement within satellite constellations.
In this section, we discuss implications of our work as well as threats to validity.

\subsection{Practical Feasibility of Satellite PoPs}

As briefly mentioned in the previous section, adding storage capabilities to satellites is a non-trivial task.
Nevertheless, such storage resources are necessary to provide satellite PoPs within a LEO constellation, as we propose with our SAT, SAT-TTL, and SAT-REP strategies.

While the idea of storage servers in satellites seems fantastical, it is feasible, as early research has shown.
Bhattacherjee et al.~\cite{Bhattacherjee2020-kr} use a Starlink satellite and commodity HPE 64-core servers to analyze the costs and challenges for such a deployment and find the results encouraging.
Neither weight, volume, nor radiation are an issue given the hardware setup and orbit altitude.
Power consumption, especially by cooling, warrants further analysis but should also not be prohibitive.
From a cost perspective, such a satellite server is estimated to be only 3x the cost of a server in a data center on earth.

Given these promising first findings and the overall low average storage requirements per satellite as determined in our evaluation, we conclude that allocating storage resources for satellite PoPs is certainly challenging yet not unreasonable.

\subsection{Impact on Request Latency}

In our evaluation, we focus on the tradeoff between two metrics: storage and bandwidth.
Request latency is a third aspect that is important for CDNs.
We omit this for two reasons:
First, latency and bandwidth are not independent of each other.
We use request hops as an estimator for bandwidth in our evaluation, but this could also be used to estimate latency.
When a request requires fewer hops, it will also incur lower latency.
Second, we have argued that one of the main advantages of satellite-backed Internet is lower request latency, as endpoints communicate over a direct connection and as ISLs benefit from a 50\% faster light propagation in a vacuum than in fiber.
For our use case, where users download web content, the latency induced by network communication even over large distances is still lower than the 80ms that humans can perceive, thus demanding no optimization~\cite{Handley2018-ay,Mohan2020-cn}.

\subsection{CDN Scalability}

Another challenge of a satellite-based CDN is scalability~\cite[p.21]{book_cloud_service_benchmarking}.
First, there is the individual PoP's scalability.
Each PoP is deployed with a fixed amount of local storage.
For both the GST and SAT strategies, this fixed amount of local storage cannot be extended as it is inaccessible, either because the equipment is managed by the end-user or because it is attached to a LEO satellite.
Compared to PoPs in terrestrial data centers, scalability could become an issue if the CDN operations show that more storage is required at specific PoPs.
The lifetime of user equipment and satellites of only a few years, however, still presents opportunities to adapt local PoP storage.

Second, there is the overall scalability of a CDN in LEO satellite constellations.
For instance, we consider the flash crowd phenomenon, or ``slashdot effect,'' where a sudden spike in popularity for a particular content occurs.
In a satellite network without integrated CDN PoPs, this would lead to all clients requesting this data from the origin servers simultaneously, a significant strain on the uplink between this server and the satellite that serves its ground terminal.
Through replication within the network, however, this bandwidth use can be limited.
The extent of this optimization depends on the number of PoPs needed to satisfy the demand for this popular data item.
With millions of ground stations fetching this item from the origin server to serve it from the local store on subsequent requests, as is the case with the GST strategy we propose, the server still has to answer millions of requests.
In the SAT, SAT-TTL, and SAT-REP strategies, however, only a handful of satellites that serve those millions of ground stations need to fetch the requested content, which is more akin to a tiered distribution.
Consequently, the load on the origin server is limited by the satellite PoPs, even in the face of this flash crowd event.
Here, the CDN leads to better scalability of the network.

And third, we must consider that a LEO satellite constellation may evolve over time, with more satellites added to increase throughput and coverage.
We have seen that the time between handoffs in the SAT-TTL and SAT-REP strategies decreases with the number of satellites, which leads to more frequent purges or more data transfer, respectively.
There may be a point at which adding additional satellites actually decreases the CDN's performance.
Depending on the size of the constellation, it can thus make sense to designate only a subset of all satellites as PoPs and take additional hops to fetch replicas over ISLs.

\subsection{Satellite Movement Prediction}

To anticipate ground station handover for the SAT-TTL and SAT-REP strategies, we use the satellite's orbital periods, which we take to be constant, and assume even spacing of satellites within a plane.
These assumptions cannot entirely hold true in the real world, as the earth is not perfectly spherical with even distribution of gravitational forces.
Gravity, atmospheric drag, and the need to dodge obstacles such as debris all influence the flight paths of satellites; their orbits and spacing may thus change over time~\cite{Blitzer1956-cz,navipedia-wu}.
To counteract this, Starlink satellites are equipped with thrusters to manipulate their flight trajectories and to de-orbit them at the end of their lifetime~\cite{spacex-js}.
We abstract from all these factors in our simulation as we have found them to only have minor impacts on the performance of the different PoP selections strategies.

\subsection{Data Consistency of Content}

A further challenge is data consistency of content~\cite{paper_bermbach_consistency}.
In our simulation, our workload is based on static content that is not updated.
In practice, content can be updated at any time, which requires invalidating local replicas at PoPs or proactively pushing replicas to these PoPs.
While the details depend on the specifics of the replication algorithms in use, we note that this is a problem that grows with the number of PoPs.
This is another disadvantage of the GST strategy compared to the SAT strategies.
Compared to the number of servers managed by Akamai, for instance, invalidating content on a few thousand satellite PoPs should be manageable.

\subsection{Dynamic Content}

When we conclude a significant improvement in bandwidth use in the network through the operation of PoPs, we can, of course, only consider content that can be replicated by CDNs.
This content is essentially static, such as images, chunked videos, scripts, or static web pages.
All other web traffic is unaffected by our optimizations as dynamically generated or user-specific content cannot be served from CDNs.
However, we observe a recent trend towards static web content that specifically aim to make additional content available through CDNs, e.g., the JAMStack~\cite{Biilmann2019-bc}.
Furthermore, Amazon Web Services already allows developers to run simple serverless functions with Lambda@Edge in some of their CloudFront PoPs, which allows dynamic content to be distributed over their CDN~\cite{Amazon_Web_Services2020-im}.
If more content providers adapt such paradigms, we could see improvements to the web's overall performance, especially in satellite networks using our proposed CDN PoP selection strategies.

\subsection{Security and Privacy}
In our paper, we did not address security aspects as data delivered by CDNs tends to be (semi-)public in practice.
If there is sensitive data, however, the GST strategy is potentially much more vulnerable as the PoP may be physically accessible.
The other three strategies open up interesting questions regarding privacy legislation since satellites fall under the jurisdiction of the state where they are registered~\cite{noauthor_1967-mk}.

%% file: Sections/6_related_work.tex
\section{Related Work}
\label{sec:relwork}

The idea of serving web content within satellite networks is not entirely new.
Gallucio et al.~\cite{Galluccio2012-ct,DOro2014-dk} present \emph{SatCache}, a scheme for content caching in information-centric satellite networks.
SatCache, however, considers only ground stations for caching.
Wu et al.\cite{Wu2016-rm} extend SatCache and develop a two-layer model that integrates caching within satellites.
They show how this second cache layer can reduce bandwidth usage when several ground stations share a single satellite.
Both papers, however, focus on static geostationary satellites rather than highly dynamic LEO constellations.

ESA's SHINE project~\cite{Romano2018-lr,shine-zv} combines satellite backhaul networks and edge distribution networks for secure distribution of multimedia content.
In the scope of the project, Luglio et al.~\cite{Luglio2018-vp} present different caching strategies and delivery models for such multimedia content.
As the project has only recently started, results are still pending.
Both SHINE and the related SCORSESE~\cite{scorsese-mq} project do not consider satellite PoPs and assume static geostationary satellites.

Liu et al.~\cite{Liu2018-xx} propose local caches in LEO satellites.
In contrast to our work, however, they present an algorithm for replicating files to satellites, while we focus on identifying the network components which shall serve as PoPs.
Furthermore, their proposed algorithm starts with a random distribution of files across all satellites before swapping files between satellites as more information about client preferences becomes available.
While their approach lowers access latency compared to SatCache, the authors ignore bandwidth impacts on ISLs.
These, however, will be severe with satellites swapping files frequently as is necessary in dynamic LEO constellations.

Both Bhosale et al.~\cite{Bhosale2020-aa} and Bhattacherjee et al.~\cite{Bhattacherjee2020-kr} propose CDN replicas in LEO satellites as potential use-cases for orbital edge computing, i.e., placing compute resources in the satellite network.
They confirm that a CDN web service is feasible on satellite servers and how it could be implemented.
We build upon this work and investigate how PoP placement for such a CDN could work.

Only a handful of research publications address the challenges of large LEO satellite networks from a perspective of Internet and network engineering:
Bhattacherjee et al.~\cite{Bhattacherjee2018-vc,Bhattacherjee2019-jz}, Klenze et al.~\cite{Klenze2018-og}, Handley~\cite{Handley2018-ay,Handley2019-ce}, Giuliari et al.~\cite{Giuliari2020-pj}, and Papa et al.~\cite{Papa2020-ny} discuss network topology and routing in LEO satellite networks.
Furthermore, Dai et al.~\cite{Dai2020-xf} emphasize the mobility of clients and user terminals and the resulting challenge of handover between access satellites.

Although content replication in satellite networks is a novel topic, data replication in dynamically changing networks in general has received widespread research attention in the context of mixed cloud/fog/edge environments~\cite{paper_bermbach_fog_vision,paper_bonomi_fog}.
For example, FBase~\cite{paper_hasenburg_towards_fbase,techreport_hasenburg_2019} proposes application-controlled replica placement in fog environments through the \emph{keygroup} abstraction that can be adapted at runtime.
This could be used to schedule data replication in accordance with the orbital mechanics of large LEO satellite constellation.
Similarly, the Global Data Plane~\cite{Zhang2015-cb}, Nebula~\cite{Ryden2014-ow}, or a combination of IPFS and RazorFS~\cite{Confais2017-bc} have been proposed as storage middleware for fog systems, yet their design makes assumptions about the underlying networks that do not hold in satellite networks.

Finally, our experiments relied on simulation as physical access to, e.g., the Starlink network, is not possible.
Here, systems such as MockFog~\cite{paper_hasenburg_mockfog,hasenburg2020mockfog} could be adapted to emulate the satellite infrastructure.
This would allow the research community to go beyond simulation and to evaluate LEO edge systems through system experiments.

%% file: Sections/7_conclusion.tex
\section{Conclusion \& Future Work}
\label{sec:conclusion}

In this paper, we have presented how the novel topology of Internet networks backed by large LEO satellite constellations are challenging the assumptions taken by CDN operators today.
Where PoPs are currently placed in Tier 3 access networks to serve groups of clients with homogenous interests and in close physical proximity, the global, converged access and backhaul network developed by companies such as SpaceX and Amazon requires re-thinking PoP placement.

We propose four strategies for PoP placement, both in ground stations and satellites.
Through simulation of a satellite constellation, we find that satellite PoPs can significantly reduce the bandwidth required to fulfill client requests without high storage requirements within the individual satellites.

The traditional CDN is not the only use-case for resource allocation in satellite networks:
with the advent of edge computing, where compute resources are available at the edge of the network, it makes sense to examine the possibilities of offloading compute tasks to Internet satellites as well.